# When the Earth and Sky Dance: Seismic Shakes Meet Weather Patterns

Alessio Kandiah*, Alexander B. Movchan*, Vladimir Frid**

*Department of Mathematical Sciences, University of Liverpool, Liverpool L69 7ZL, UK

**Sami Shamoon College of Engineering, Jabotinsky 84, 77245 Ashdod, Israel

**Abstract.** A new modelling approach shows how the Earth's hidden vibrations may drive global weather dynamics and atmospheric pressure variations, hinting that the planet's own beat could be imprinted on our climate. The atmospheric rotational patterns of the mean sea level pressure, in connection to the development of powerful storms, are shown to be caused by Earth's rotational elastic dynamics and earthquake-induced oscillations. These seismic excitations are discussed in relation to storm formation and the global atmospheric patterns of high-pressure regions.

**Introduction.** The years 2024-2025 were marked by a series of extreme seismic events and unusually high variations of mean sea pressure level - from as low as 921 hPa to as high as 1060 hPa, which was similar to the years 1883-1884, with one of the largest eruptions of Krakatoa in 1883, and the lowest ever recorded mean sea level pressure in the British Isles of 926 hPa in January of 1884, as described in the 1884 and 1930 Nature articles[1,2]. Recent research[3,4,5] reveals that our planet undergoes elastic oscillations[3], resonating like an immense gyroscope. The natural vibrations of the Earth ripple through the tectonic plate boundaries and echo in the atmosphere[6], revealing hidden connections between the Earth's elastic dynamics and changing weather patterns. By analysing the natural resonances, scientists can uncover how the planet's movement may align with atmospheric pressure variations, providing a new perspective on the interconnected rhythms of the Earth and sky. Numerical models tie the Earth's vibrations to real-world atmospheric variability, suggesting that the planet's 'heartbeat' may be felt both beneath our feet and above our heads.

A striking characteristic of planetary atmospheres in the polar regions is the polygonal-like structures of fast-moving air high in the atmosphere, shaped by the combined effects of the planet's rotation and gravity. As shown in Fig. 1(a), Saturn's North Pole hosts a persistent hexagonal jet stream, first detected by Voyager and later confirmed by Cassini NASA probes. A similar phenomenon appears on Earth in the form of an approximate pentagonal jet-stream pattern at the South Pole as illustrated in Fig. 1(b). These polygonal formations highlight how physical mechanisms, such as gyroscopic effects and gravitational forces, can produce highly ordered atmospheric structures.

A new perspective is presented here, following the idea that atmospheric patterns may also be influenced by the Earth's elastic vibrations[6]. By examining mean sea-level pressure variations during recent extreme events, the analysis reveals how Earth's hidden vibrational modes may couple with the atmosphere to shape large-scale weather dynamics.

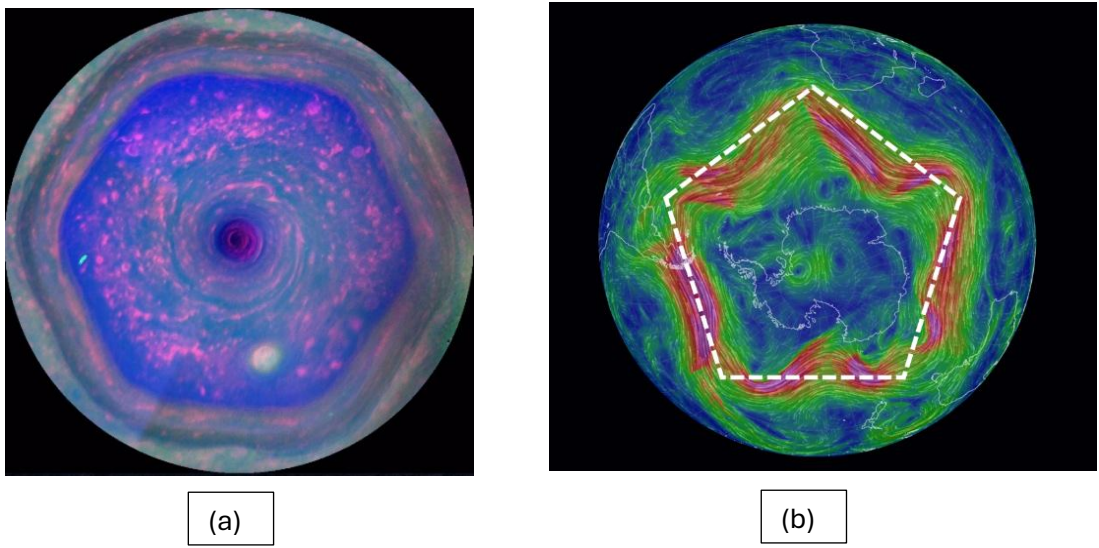

(a)  (b)

Figure 1: (a) Six-sided jet stream at Saturn's North Pole from NASA's Cassini mission. Image courtesy of NASA as of December 2025, and (b) approximate pentagonal jet stream at the South Pole of Earth on 3rd January 2023 (produced using the dataset Earth_pentagon).

**How Earth's rotation shapes its vibrations.** It is known that the Earth vibrates with natural oscillations governed by elasticity, fluid dynamics and gravity[7,8]. For a non-rotating planet, these vibrations settle into the familiar toroidal and spheroidal modes. But once rotation is taken into account, the Coriolis effect transforms the picture: the vibration modes engage a coupled oscillatory system with striking new patterns that would not exist in a non-rotating planet. These rotational effects shift the eigenfrequencies and change eigenmodes, leading to phenomena such as gyroscopic frequency splitting and the Chandler wobble[9]. The study[3] explores different classes of vibrations of rotating elastic bodies, with physical scales that approximate Earth itself. The results revealed approximately polyhedral vibration modes in a spinning Earth-like ball, uncovering hidden resonances driven by the planet's spin.

At low frequencies, the rotating elastic ball reveals distinct bands that run parallel to the equator[3,4,5]. These patterns emerge since rotation induces anisotropy through the Coriolis effect, structuring the displacement field of the planet. The bands mark regions of relatively low-amplitude elastic motions, separated by boundaries where oscillations intensify. The results of the numerical computations are shown in Fig. 2 for two different vibration frequencies. On Earth, comparable banded structures appear in atmospheric circulation patterns, known as the Hadley, Ferrel and Polar cells that organise global weather systems into repeating latitudinal patterns.

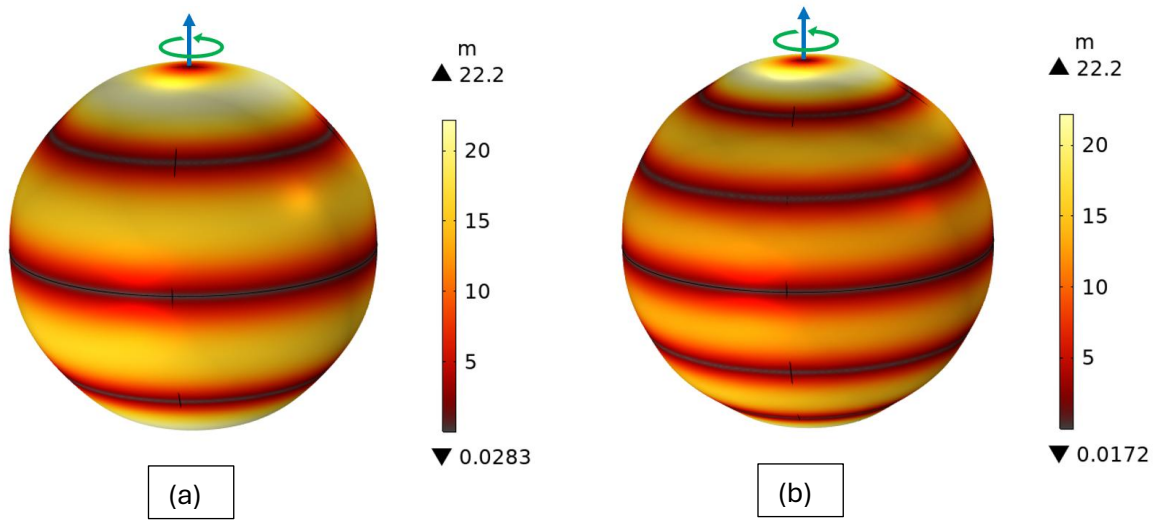

Figure 2: Axially symmetric eigenmodes of a rotating elastic ball, showing low-amplitude displacement bands parallel to the equatorial region for two vibration frequencies: (a) $f = 3.5188 \times 10^{-4}$ Hz, and (b) $f = 5.1216 \times 10^{-4}$ Hz. The material properties are: Young's modulus 110 GPa, Poisson's ratio 0.3 and mass density 5515 kg/m³. The ball rotates about its vertical axis, passing through the centre and poles, with an angular speed of $7.2921 \times 10^{-5}$ rad/s.

At certain frequencies, the rotating elastic ball's vibrations may also resemble icosahedral or dodecahedral structures illustrated in Fig. 3, with pentagonal or triangular patterns across the poles and equatorial regions. These polyhedral eigenmodes depend on both the rotational speed and material properties of the rotating body.

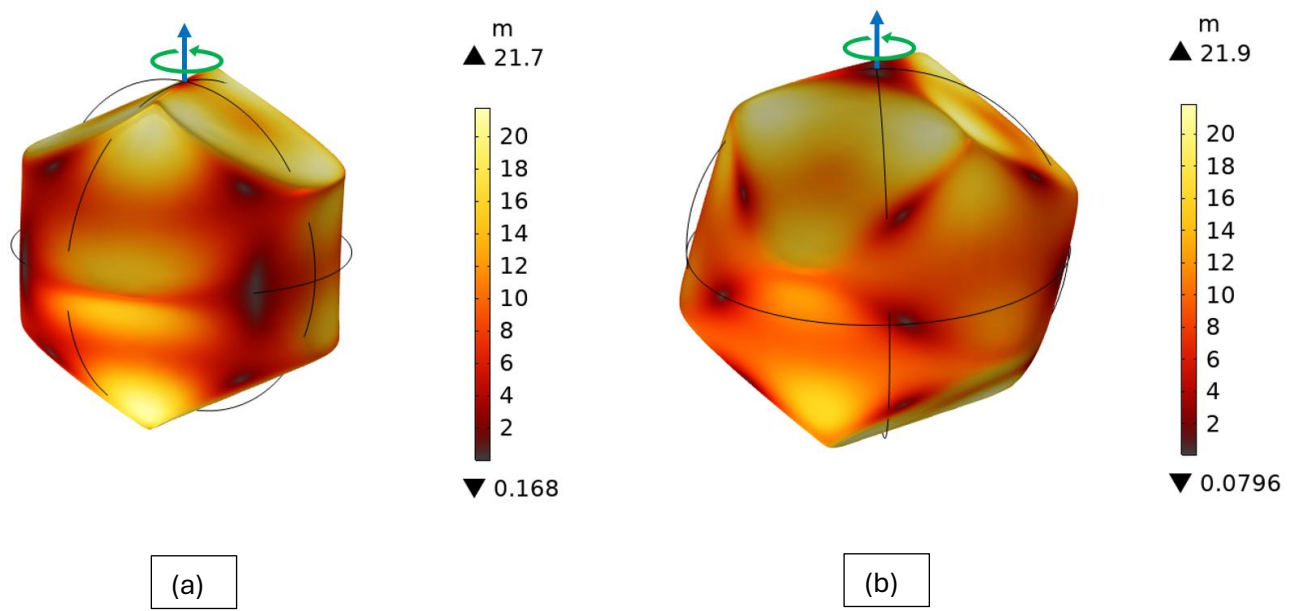

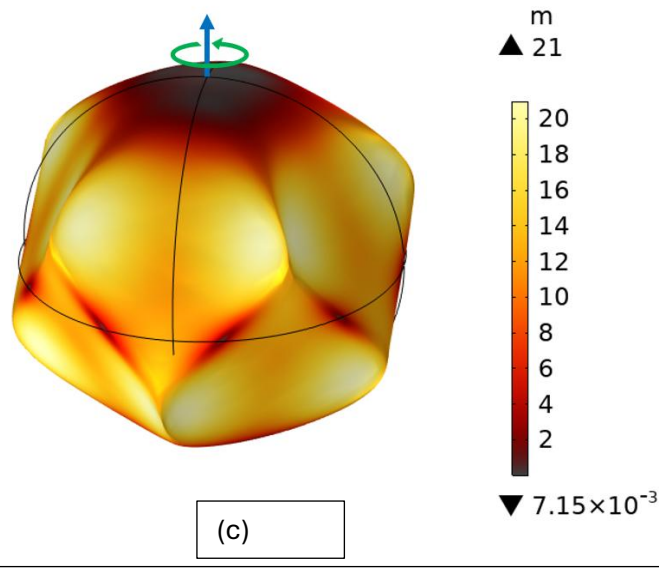

(c)

Figure 3: Polyhedral vibration modes of a rotating isotropic elastic ball, with rotational symmetry about the vertical $z$-axis. The material properties and the size of the ball are identical to those in Fig. 2. (a) Eigenmode with two-fold rotational symmetry for the oscillation frequency $f = 7.8146 \times 10^{-4}$ Hz, (b) eigenmode with three-fold rotational symmetry for the vibration frequency $f = 7.8420 \times 10^{-4}$ Hz, and (c) eigenmode with five-fold rotational symmetry for the oscillation frequency $f = 7.8483 \times 10^{-4}$ Hz.

The study of wave phenomena on spherical geometries reveals that the Earth's rotation results in elastic vibrations with remarkable patterns. Computations in Fig. 3 show striking vibration modes with two-fold, three-fold and five-fold rotational symmetries, connected to global high-pressure regions experienced during high-intensity storms and earthquakes[3]. Just as storms and high-pressure regions organise into repeating patterns around the globe, the planet's vibrations exhibit rotational two-fold, three-fold or five-fold patterns, establishing the coupling between Earth's elastic dynamics and atmospheric variability.

**Earth's vibrations echoing in the atmosphere: recent global storms and pressure variations.** Between October 2024 and November 2025, clusters of seismic activity across the globe occurred alongside extraordinary atmospheric disturbances and variations in pressure levels. The sequence of recent events, including typhoon Ragasa, storms Éowyn, Amy, Floris and Benjamin, and the resulting five-fold rotational patterns of high *Mean Sea Level Pressure* reveal how Earth's rotational dynamics, elastic vibrations and atmospheric pressure gradients combine to produce large-scale weather systems.

**Typhoon Ragasa** (17–25 September 2025) hit the Philippines, Taiwan and China with maximum recorded winds reaching 270 km/h and extreme pressure variations. There was no advance warning. However, there was a series of significant seismic events during 17-19 September 2025, with the strongest earthquake of Mag 7.8 in Kamchatka on 18 September 2025. According to the meteorological data, shown in Fig. 4, a five-fold rotational pattern of high-pressure regions was formed in the atmosphere. On 22nd September 2025, the recorded mean sea level pressure was 928 hPa in the vicinity of the Philippines as shown in Fig. 4(b), which also shows the approximate pentagonal high-pressure structure. In particular, Fig. 4(c) shows that the vortex signifying Ragasa was formed as a result of the interaction between two closely located high-pressure systems.

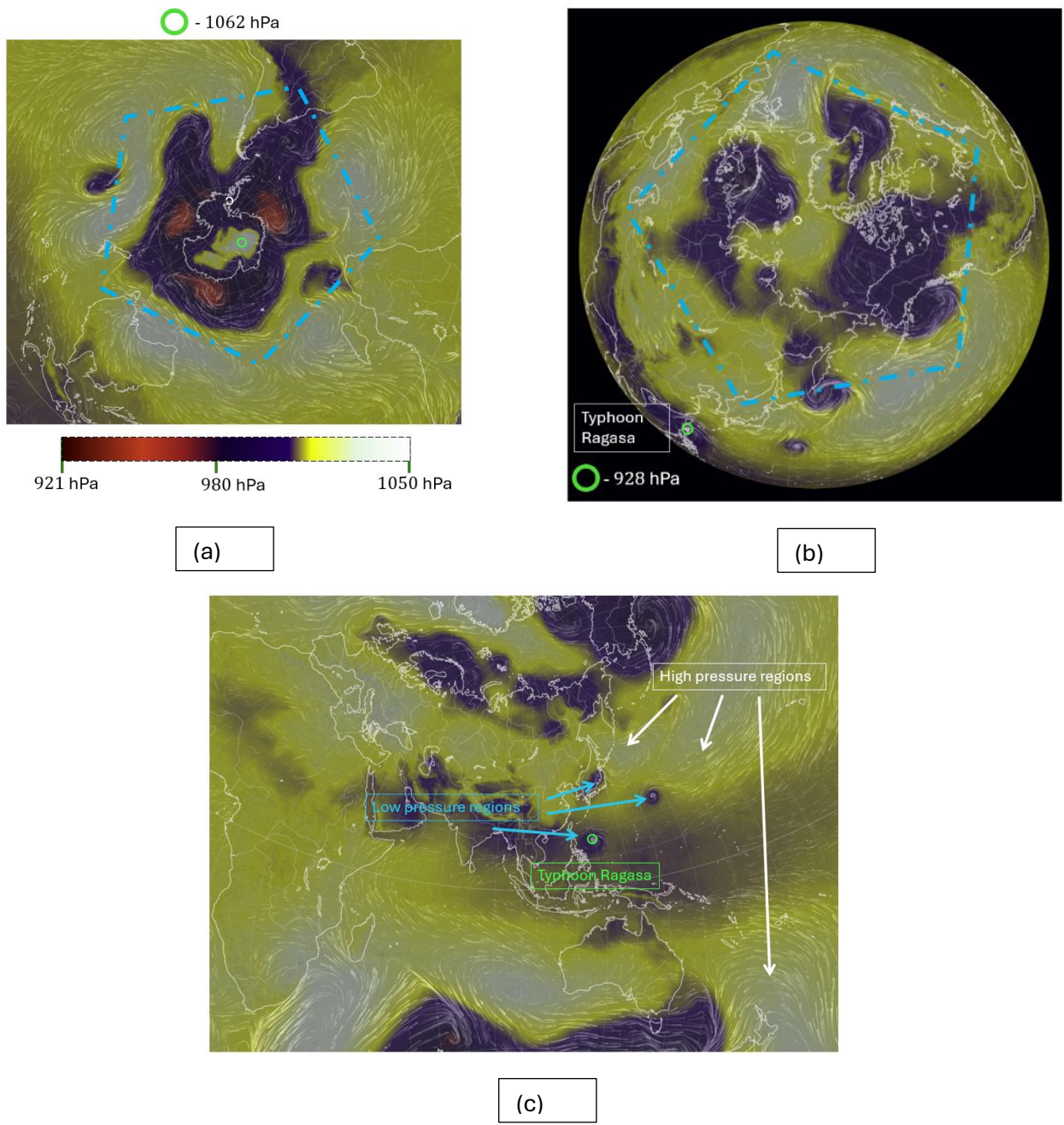

Figure 4: Mean sea level pressure and surface wind across the Earth's surface prior and during Typhoon Ragasa. (a,b) Five-fold rotational patterns of high pressure are shown at the South and North Poles, produced using the datasets Earth_S and Earth_N. (c) Formation of low pressure vortices (including Ragasa) induced by the interaction between the high pressure systems on 20th September 2025, produced from the dataset Ragasa_20.

**Storm Éowyn** (21-27 January 2025) swept across Ireland and the United Kingdom, driven by extreme pressure gradients. On 24th January 2025, the mean sea level pressure varied significantly between 942 hPa and 1060 hPa within the region shown in Fig. 5(a). These variations were accompanied by approximately pentagonal patterns with localised high-pressure regions in the atmosphere. Bearing in mind that extreme pressure changes were the dominant characteristics of the Storm Éowyn, there were a cluster of earthquakes prior to this storm, which included a Mag 7.1 earthquake in Tibet, China on 7th January 2025, a Mag 6.8 earthquake in Kyushu, Japan on 13th January 2025 and a Mag 6.0 earthquake in Yujing, Taiwan on 20th January 2025. Interestingly, similar climatic conditions were observed in January 1884, when strong continuous seismic activity preceded record-low pressures of 926.5 hPa in Scotland[1,2]. It is also worthwhile noting that in the first week of February 2025 (few days after Storm Éowyn), the Mean Sea Level pressure in the UK increased to the unusually high level of 1045hPa (the highest Mean Sea Level Pressure recorded in the UK was 1054 hPa in 1902).

Later, **Storm Amy** (1–6 October 2025) tracked northeast across the UK, causing severe disruption and displaying a five-fold rotational pattern of high-pressure regions as shown in Fig. 5(b). The following earthquakes preceded Storm Amy: Mag 7.4 and Mag 7.8 earthquakes in Kamchatka, Russia on 13th and 18th September 2025, respectively, Mag 6.2 and Mag 6.3 earthquakes in Zulia, Venezuela on 24th and 25th September 2025, respectively, and a Mag 6.9 earthquake in Leyte, Philippines on 30th September 2025.

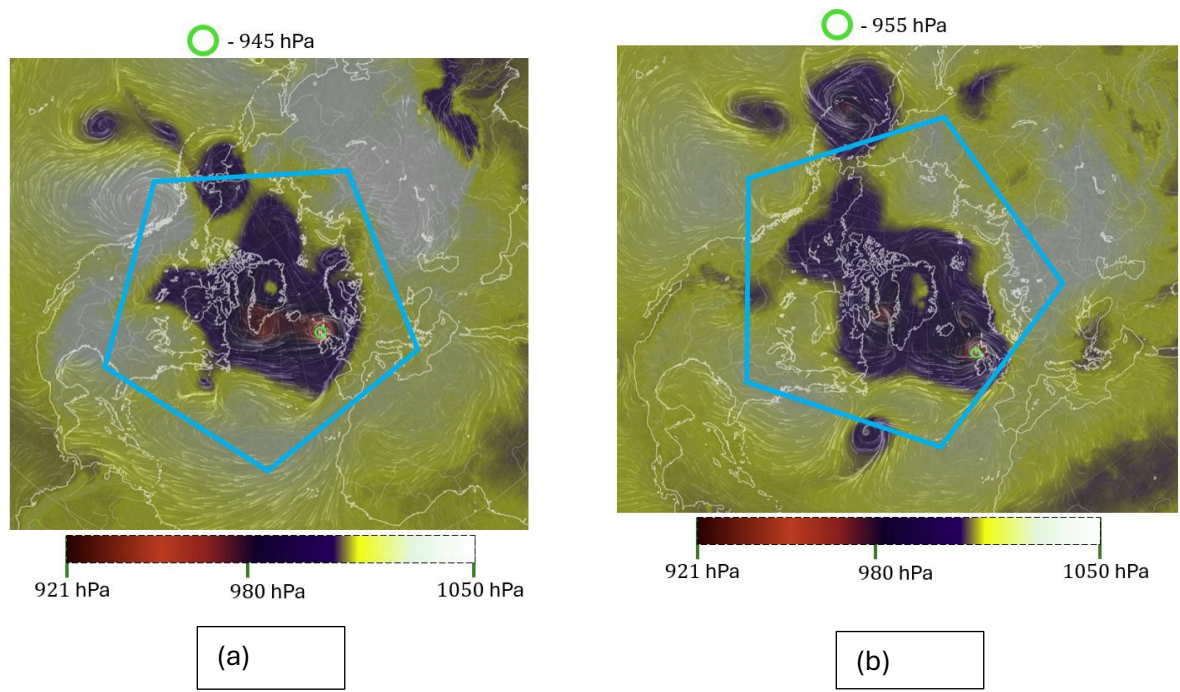

Figure 5: Stereographic projections of Earth's surface winds and mean sea level pressures during (a) Storm Éowyn (produced from dataset Eowyn) and (b) Storm Amy (produced from dataset Amy). The examples also highlight the approximate five-fold symmetry patterns of the localised high-pressure systems, influenced by the vibrations of the planet and changing pressure gradients.

Remarkably, **Storm Floris** (2-4 August 2025) and **Storm Benjamin** (22-23 October 2025) were also linked to localised high-pressure systems as shown in Fig. 6, which were preceded by strong seismic activity. Storm Floris (2–4 August 2025) brought strong winds and heavy rain in northern Scotland, with an approximate three-fold pattern of high-pressure regions as illustrated in Fig. 6(a). Storm Benjamin (22–23 October 2025), that affected the UK, was surrounded by localised high-pressure regions with an approximately two-fold pattern, distinct from earlier storms, as shown in Fig. 6(b). The systems of high-pressure, discussed above, resemble the patterns produced by the vibration modes of the rotating elastic ball presented in Fig. 3. This is consistent with the observation that the Earth's rotational dynamics and elastic vibrations, which include seismic excitations[10], may couple with atmospheric circulation[6].

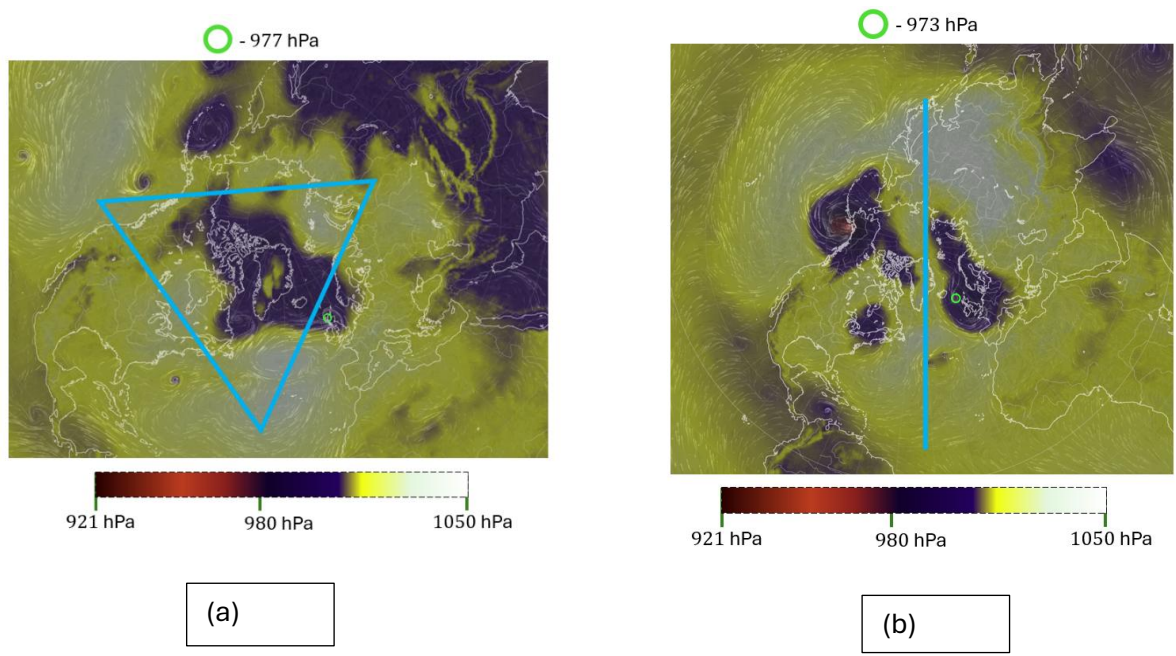

(a)  (b)

Figure 6: Stereographic projections illustrating Earth's surface winds and mean sea level pressures during (a) Storm Floris (produced from dataset Floris) and (b) Storm Benjamin (produced from dataset Benjamin). The examples also demonstrate the symmetry of the localised high-pressure regions in connection with the elastic vibrations of the Earth.

**Concluding remarks.** These storms highlight how the elastic oscillations of the rotating Earth resonate with atmospheric circulation, producing localised high pressure patterns, and vortices. Planetary vibrations together with atmospheric pressure variations reveal the unity between Earth's elastic rotational dynamics, vibroseismological processes and weather systems on a global scale.

# References

1. The Low Barometer of January 26, 1884. Nature **30**, 58-59 (1884). https://doi.org/10.1038/030058a0
2. Historic Natural Events. Nature **125**, 149 (1930). https://doi.org/10.1038/125149a0
3. Barnfield, W. B., Kandiah, A., Frid, V., Movchan, I. B. & Movchan A. B. Global lattice models of a rotating planet-view on shake, rattle and roll. Preprint at ArXiv https://doi.org/10.48550/arXiv.2508.16630 (2025)
4. Kandiah, A., Jones, I. S., Movchan, N. V. & Movchan A. B. Dispersion and asymmetry of chiral gravitational waves in gyroscopic mechanical systems. Part 1: Discrete lattice strips. QJMAM. **78**, hbaf004 (2025)
5. Kandiah, A. Jones, I. S., Movchan, N. V. & Movchan A. B. Dispersion and asymmetry of chiral gravitational waves in gyroscopic mechanical systems. Part 2: Continuum asymptotic models in equatorial and polar regions. QJMAM. **78**, hbaf005 (2025)
6. Carbone, V. et al. A mathematical model of lithosphere-atmosphere coupling of seismic events. Sci. Rep. **11**, 8682 (2021)
7. Suda, N., Nawa, K. & Fukao, Y. Earth's background free oscillations. Science. **279**, 2089-2091 (1998)
8. Alterman, Z., Jarosch, H. & Pekeris, C. L. Oscillations of the Earth. Proc. R. Soc. A. **252**, 80-95 (1959)
9. Montagner, J. & Roult, G. Normal modes of the Earth. Phys. Conf. Ser. **118**, 012004 (2008)
10. Benioff, H., Press, F. & Smith, S. Excitation of the free oscillations of the Earth by earthquakes. Geophys. Res. **66**, 605-619 (1961)

## Acknowledgements

A.K. gratefully acknowledges the financial support of the EPSRC through the Mathematics DTP grant EP/V52007X/1, project reference 2599756. A.B.M. and V.F. are grateful to the Wohl Clean Growth Alliance of the British Council, for supporting the academic visit of V.F. to the University of Liverpool in 2024 and 2025, and the visit of A.B.M. to Sami Shamoon College of Engineering in 2025.

## Competing interests

The authors declare no competing interests.